\documentstyle[preprint,aps,epsf]{revtex}

\newcommand{\bd}{\begin{document}}
\newcommand{\ed}{\end{document}}
\newcommand{\bc}{\begin{center}}
\newcommand{\ec}{\end{center}}
\newcommand{\be}{\begin{eqnarray}}
\newcommand{\ee}{\end{eqnarray}}
\renewcommand{\thefootnote}{\alph{footnote}}
\newcommand{\se}{\section}
\newcommand{\sse}{\subsection}
\newcommand{\bi}{\bibitem}
\def\figcap{\section*{Figure Captions\markboth
     {FIGURECAPTIONS}{FIGURECAPTIONS}}\list
     {Figure \arabic{enumi}:\hfill}{\settowidth\labelwidth{Figure 999:}
     \leftmargin\labelwidth
     \advance\leftmargin\labelsep\usecounter{enumi}}}
\let\endfigcap\endlist \relax
\input psfig

\begin{document}
 \pagestyle{empty} \draft
\vfill
\title{
Long distance contributions in $B\to K^* \ell^+ \ell^-$
decays with~polarized~$K^*$
}
\author{Chuan-Hung Chen$^a$ and C.~Q.~Geng$^{b,c}$}
 \address{  \sl ${}^a$Institute of Physics, Academia Sinica, Taipei,
Taiwan 115, ROC }
 \address{\sl ${}^b$Department of Physics, National Tsing Hua University,
Hsinchu, Taiwan 300, ROC }
 \address{\sl $^c$Theory Group, TRIUMF, 4004 Wesbrook Mall, Vancouver,
  B.C. V6T 2A3, Canada}
\vfill
\maketitle
\begin{abstract}

We use momentum correlations as physical observables in $B\to
K^{*}l^{+}l^{-}$ decays with $K^{*}$ polarized to study the long
distance contributions. We show that these observables are
sensitive to the scenarios of the long distance parametrizations.
We find that the T-odd observable is directly related to the
nonfactorizable effect in the standard model.

\end{abstract}
%
\pagestyle{plain}

The study of flavor changing neutral currents (FCNCs) in B decays
has an enormous progress since the CLEO observation \cite{cleo} of
 $b\to s\gamma $.
Recently, the process of $B\to Kl^+l^-$ has been also observed
\cite{Belle} at the Belle detector in the KEKB $e^+e^-$ storage
ring. It is known that the radiative $b\to s\gamma $ and
semileptonic $b\to s l^+l^-$ FCNC decays \cite{Bdecays} in the
standard model (SM) provide us with information on not only the
 Cabibbo-Kobayashi-Maskawa (CKM) matrix elements
\cite{ckm}  but also  physics beyond the SM. Moreover,
for $b\rightarrow s l^+ l^-$, new operators such as those from the
box and Z-penguin diagrams can escape the strict constraint from
$b\rightarrow s \gamma$ and, therefore, the new physics effect
could be sizable.

In addition to the short-distance (SD) contributions, the
long-distant (LD) contributions to $b\to s\gamma \ (sl^{+}l^{-})$,
arising from the charm ($c $) quark pair bound states, should be
taken into account. It is known that the LD effect in
$b\rightarrow s\gamma $ is only a few percent and negligible,
whereas it is the main part to the decay rate in $b\rightarrow
sl^+l^-$. However, the parametrization of the LD contributions is
not unique and has an
uncertainty of about $20\%$ for the decay branching ratios (BRs) of $%
b\rightarrow sl^+l^-$ \cite{Ahmady}. In order to test the SM and
find new physics,
it is of important to extract such theoretical uncertainty. To
distinguish  various theoretical parametrizations, it is
interesting to see if we can find some measurable physical observables
which are dominated by the LD parts.

In this paper, we will study
the LD effects by considering the exclusive $B\rightarrow
K^{*}l^{+}l^{-}$ decays with the polarized $K^{*}$ meson.
We will define some useful observables by the momentum
correlations, especially those related to T-odd operators. In a
three-body decay, it is known that the simplest T-odd operator is
the triple correlations given by $\vec{s}\cdot ( \vec{p}_{i}\times
\vec{p}_{j})$
where $\vec{s}$ is the spin vector of an outgoing particle and $\vec{p}%
_{i}$ and $\vec{p}_{j}$ denote any two independent momentum
vectors. In terms of the CPT invariant theorem, T violation (TV)
implies CP violation (CPV). Therefore, studying of T-odd
observables could help us to understand the origin of CPV. We note
that the T-odd observables such as the triple correlations
are only associated with the imaginary parts of relevant dynamical
variables. That is, even there is no weak CP phase, these
observables may not vanish if a strong phase or absorptive part
exists. In the SM, since the CKM matrix element of
$V_{tb}V_{ts}^{*}$ involved in the process of $B\rightarrow
K^{*}l^{+}l^{-}$ contains no phase, the T-odd observables can be
only generated through the LD effects.
  Hence, these observables can be used
 to test the parametrizations of LD effects. In the
decays of $B\rightarrow K^{*}l^{+}l^{-}$ ($l=e,\mu $, and $\tau
$), the spin $s$ can be the polarized lepton, $s_{l},$ or the
$K^{*}$ meson, $\epsilon ^{*}(\lambda )$. For the polarized
lepton,
since the T-odd transverse lepton
polarization flips the helicity and thus it is always associated with the
lepton mass,
we expect that this type of T violating effects is suppressed and less
than $1\%$ for the light lepton modes \cite {CG1}.
Such effect is also negligible for the $\tau $ mode due to the
small decay branching ratio.
In this paper, we will concentrate on the light lepton modes with
only $K^*$ polarized and set the lepton masses to be zero, i.e.,
$m_{l}=0$.

We start by writing the effective Hamiltonian for $b\rightarrow s\ l^{+}\
l^{-}$ as \cite{Buras}
\begin{equation}
{\cal H}=\frac{G_{F}\alpha V_{tb}V_{ts}^{*}}{\sqrt{2}\pi }\left[ H_{1\mu
}L^{\mu }+H_{2\mu }L^{5\mu }\right]  \label{heff}
\end{equation}
with
\begin{eqnarray*}
H_{1\mu } &=&C_{9}(\mu )\bar{s}\gamma _{\mu }(\mu )P_{L}b\ -\frac{2m_{b}}{%
q^{2}}C_{7}(\mu )\bar{s}i\sigma _{\mu \nu }q^{\nu }P_{R}b \,,\\
H_{2\mu } &=&C_{10}\bar{s}\gamma _{\mu }P_{L}b \,,\ \
L^{\mu } \;=\;\bar{l}\gamma ^{\mu }l\,,\ \
L^{5\mu } \;=\;\bar{l}\gamma ^{\mu }\gamma _{5}l\,,
\end{eqnarray*}
where  $C_{i}\ (i=7,9,10)$ are
the Wilson coefficients (WCs) and their expressions can be found in Ref.
\cite{Buras} for the SM. Since the operator associated with $C_{10}$ is not
renormalized under the QCD, it does not depend on the renormalization
scale.
As mentioned before, besides the short-distance (SD) contributions, the main
effect on the BR comes from c\={c} resonant states such as $\Psi$ and
$\Psi^{\prime }$. In the literature \cite{Ahmady,DTP,LMS,AMM,OT,KS},
it has been suggested to combine the factorization assumption (FA) and
vector meson dominance (VMD) approximation in estimating LD effects.
 As a consequence, these effects can be absorbed to the
relevant WC of $C_{9}$. For comparing the different
parametrizations, we adopt three scenarios in the
literature for the effective WC of $C_{9}$:

(I) By defining
\begin{equation}
\left\langle 0\right| \bar{c}\gamma _{\mu }c\left| V(q)\right\rangle
=\varepsilon _{\mu }f_{V}(q^{2})\,,
\end{equation}
where $\varepsilon _{\mu }$ denotes the polarization vector of $V$,
and fixing $f_{V}(q^{2})$ at the $V$ mass-shell with $q^{2}=m_{V}^{2}$,
one has that
\begin{equation}
C_{9}^{eff}=C_{9}\left( \mu \right) +\left( 3C_{1}\left( \mu \right)
+C_{2}\left( \mu \right) \right) \left( h\left( x,s\right) -\frac{3}{\alpha
^{2}}\sum_{V=\Psi ,\Psi ^{\prime }}k_{V}\frac{\pi \Gamma \left( V\rightarrow
l^{+}l^{-}\right) M_{V}}{M_{V}^{2}-q^{2}-iM_{V}\Gamma _{V}}\right) \,,
\label{effc91}
\end{equation}
where $h(x,s)$ describes the one-loop matrix elements of operators
$O_{1}= \bar{s}_{\alpha }\gamma ^{\mu }P_{L}b_{\beta }\
\bar{c}_{\beta }\gamma _{\mu }P_{L}c_{\alpha }$ and
$O_{2}=\bar{s}\gamma ^{\mu }P_{L}b\ \bar{c}\gamma _{\mu }P_{L}c$
\cite{Buras}, $M_{V}$ ($\Gamma _{V}$) are the masses (widths) of
intermediate states, and the factors $k_{V}\approx 2.3$ are
phenomenological parameters for compensating the approximations of
the FA and VMD and reproducing the correct branching ratios $Br\left(
B\rightarrow J/\Psi X\rightarrow l^{+}l^{-}X\right) =Br\left(
B\rightarrow J/\Psi X\right) $ $\times $ $Br\left( J/\Psi
\rightarrow l^{+}l^{-}\right) $. Here, we have neglected the small
Wilson coefficients.

(II) By parametrizing
$f_{V}(q^{2})$ as \cite{Ahmady}
\begin{equation}
f_{V}(q^{2})=f_{V}(0)\left( 1+\frac{q^{2}}{c_{V}}\left( d_{V}-h_{V}\left(
q^{2}\right) \right) \right)
\end{equation}
where $c_{\Psi (\Psi ^{\prime })}=0.54$ $(0.77)$, $d_{\Psi (\Psi ^{\prime
})}=0.043$ and
\[
h_{V}\left( q^{2}\right) =\frac{1}{16\pi ^{2}r}\left[ -4-\frac{20r}{3}%
+4\left( 1+2r\right) \sqrt{1-\frac{1}{r}}\arctan \frac{1}{\sqrt{1-\frac{1}{r}%
}}\right]
\]
with $r\approx q^{2}/m_{V}^{2}$ for $0\leq q^{2}\leq m_{V}^{2}$ and $%
f_{V}(q^{2})=f_{V}(m_{V}^{2})$ for $q^{2}>m_{V}^{2}$, one gets that
\begin{equation}
C_{9}^{eff}=C_{9}\left( \mu \right) +\left( 3C_{1}\left( \mu \right)
+C_{2}\left( \mu \right) \right) \left( h\left( x,s\right) -\frac{3}{\alpha
^{2}}\sum_{V=\Psi ,\Psi ^{\prime }}\frac{f_{V}^{2}(q^{2})}{%
f_{V}^{2}(m_{V}^{2})}\frac{\pi \Gamma \left( V\rightarrow l^{+}l^{-}\right)
M_{V}}{q^{2}-M_{V}^{2}-iM_{V}\Gamma _{V}}\right) \,.  \label{effc92}
\end{equation}

(III) With the measurement of $R_{had}(q^{2})\equiv \sigma
(e^{+}e^{-}\rightarrow hadron)/\sigma (e^{+}e^{-}\rightarrow \mu ^{+}\mu
^{-})$ and the dispersion relation \cite{KS},
one finds that
\begin{equation}
C_{9}^{eff}=C_{9}\left( \mu \right) +Y^{\prime}(s)+\left( 3C_{1}\left( \mu
\right) +C_{2}\left( \mu \right) \right) \left( {Re}\,g\left( \hat{m}%
_{c},s\right) +i{Im}\,g\left( \hat{m}_{c},s\right) \right)  \label{effc93}
\end{equation}
where $\hat{m}_{c}=m_{c}/m_{b}$, $s=q^{2}/m_{b}^{2}$, $Y^{\prime}(s)$
is defined in Ref. \cite{Buras}, and
\begin{eqnarray*}
{Re}\,g\left( \hat{m}_{c},s\right) &=&-\frac{8}{9}\ln \hat{m}_{c}-\frac{4}{9}%
+\frac{s}{3}P\int_{4\hat{m}_{\pi }^{2}}^{\infty }ds^{\prime }\frac{R_{had}^{c%
\bar{c}}\left( s^{\prime }\right) }{s^{\prime }\left( s^{\prime }-s\right) },
\\
{Im}\,g\left( \hat{m}_{c},s\right) &=&\frac{\pi }{3}R_{had}^{c\bar{c}}\left(
s\right) ,\ \
R_{had}^{c\bar{c}}\left( s\right) \;=\;R_{cont}^{c\bar{c}}\left( s\right)
+R_{res}^{c\bar{c}}\left( s\right)\,,
\end{eqnarray*}
where
$P$ denotes the principal value and $R_{cont}^{c\bar{c}}\left( s\right) $
and $R_{res}^{c\bar{c}}\left( s\right) $ are the contributions of continuum
and resonant states with the explicit expressions given by
\begin{eqnarray*}
R_{cont}^{c\bar{c}}\left( s\right) &=&\left\{
\begin{tabular}{lll}
$0$ &  & for $0\leq s\leq 0.60$ \\
$-6.8+11.33s$ &  & for $0.60\leq s\leq 0.69$ \\
$1.02$ &  & for $0.69\leq s\leq 1,$%
\end{tabular}
\right. \\
R_{res}^{c\bar{c}}\left( s\right) &=&\frac{9q^{2}}{\alpha ^{2}}\sum_{V=\Psi
,\Psi ^{\prime }}k_{V}\frac{Br\left( V\rightarrow l^{+}l^{-}\right) \Gamma
_{V}\Gamma _{had}^{V}}{\left( q^{2}-M_{V}^{2}\right) ^{2}+M_{V}^{2}\Gamma
_{V}^{2}}\,.
\end{eqnarray*}

\begin{figure}[tbp]
\centerline{ \psfig{figure=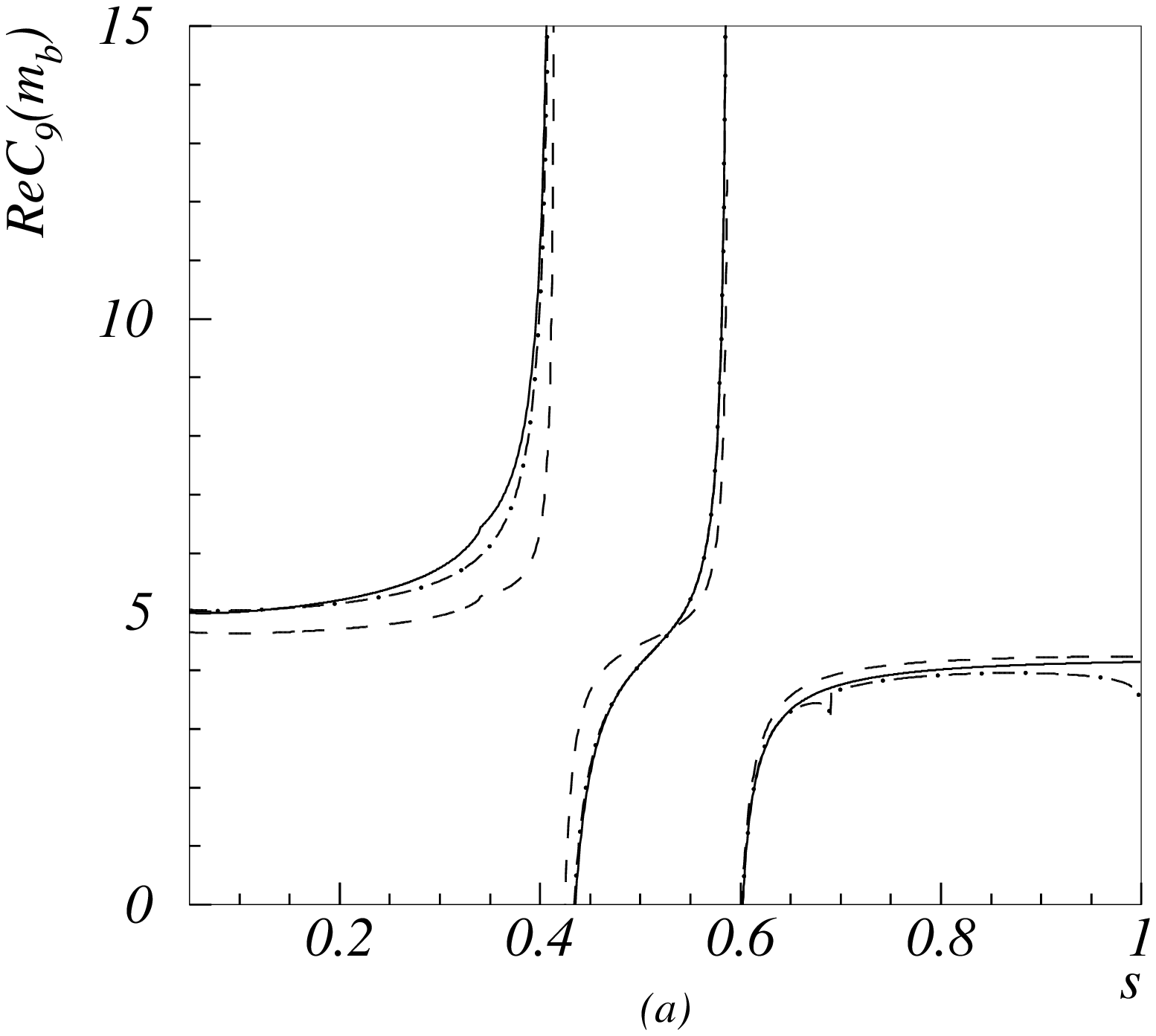,height=2.4in }$\ \
$\psfig{figure=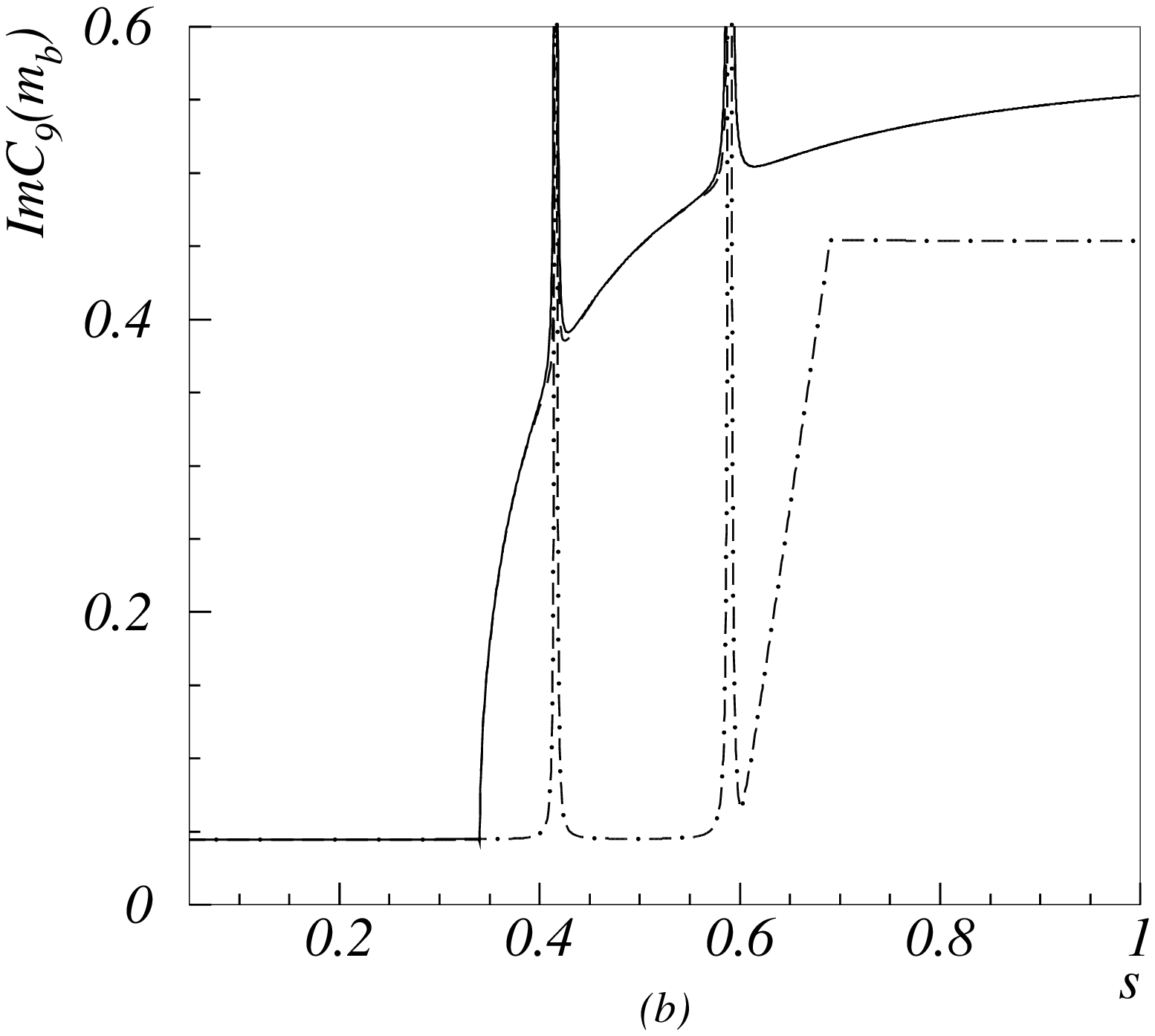,height=2.4in }} \caption{Effective WCs of
(a) $ReC^{eff}_{9}$ and (b) $ImC^{eff}_{9}$. The solid, dashed and
dash-dotted lines correspond to the scenarios of (I), (II) and
(III), respectively.} \label{f1}
\end{figure}

 In Figure \ref{f1}, we plot the real and imaginary parts of $%
C^{eff}_{9}$ for the three scenarios. From the figure, we clearly see that
the results for $ReC^{eff}_{9}$ in (I) and (III) are close to each other and
slightly different from that in (II), whereas that for $ImC^{eff}_{9}$ in
(I) and (II) are almost the same but quite different from (III).

In addition, we note that
the LD contributions to $
BR(B\rightarrow K^{*}\gamma )$ are pure nonfactorizable effects
and only at a few percent level \cite{KRSW},
whereas they are enormous around $c\bar{c}$ resonant
states for $B\rightarrow K^{*}l^{+}l^{-}$.
 From Ref. \cite{MNS}, similar to the factorizable effects  to
 $C_{9}$, the nonfactorizable contributions to $b\rightarrow s
\gamma$ can be put into $C_{7}$, given by
\begin{equation}
C_{7}^{eff}=C_{7}\left( \mu \right) +\omega \Delta C_{9}^{eff}\left( \mu
\right)  \label{effc7}
\end{equation}
with $\Delta C_{9}^{eff}\left( \mu \right) =C_{9}^{eff}\left( \mu \right)
-C_{9}\left( \mu \right) $, where $\omega $ parametrizes the magnitude of
the ratio of nonfactorizable and factorizable parts.
By satisfying the present
experimental constraint on $BR(B\rightarrow K^{*}\gamma )$ at $q^{2}=0$,
we set $\omega \leq 0.15$. If the $\omega $ effect is displayed
exclusively, we can directly demonstrate the magnitude of
nonfactorizable effects.
We also note that nonfactorizable effects in $B\to K^*$ decays have been
computed systematically in the QCD factorization approach \cite{BF}.

For $B\to K^{*}l^{+}l^{-}$ decays, the relevant transition form factors can
be parametrized as
\begin{eqnarray}
< K^{*}(p_{2},\epsilon )| \bar{s}\gamma _{\mu }b| \bar{B}(p_{1})>
&=&iV(q^{2})\varepsilon _{\mu \alpha \beta \rho }\epsilon ^{*\alpha
}P^{\beta }q^{\rho },  \nonumber \\
<K^{*}(p_{2},\epsilon )| \bar{s}\gamma _{\mu }\gamma _{5}b| \bar{B}(p_{1})>
&=&A_{0}(q^{2})\epsilon _{\mu }^{*}+\epsilon ^{*}\cdot q\left(
A_{1}(q^{2})P_{\mu }+A_{2}(q^{2})q_{\mu }\right) ,  \nonumber \\
<K^{*}(p_{2},\epsilon )| \bar{s}i\sigma _{\mu \nu }q^{\nu}b| \bar{B}(p_{1})>
&=&iT(q^{2})\varepsilon _{\mu \alpha \beta \rho }\epsilon ^{*\alpha
}P^{\beta }q^{\rho },  \nonumber \\
< K^{*}(p_{2},\epsilon )| \bar{s}i\sigma _{\mu \nu }q^{\nu}\gamma _{5}b|
\bar{B}(p_{1})> &=&-T_{0}(q^{2})\epsilon _{\mu }^{*}-\epsilon ^{*}\cdot
q\left( T_{1}(q^{2})P_{\mu }+T_{2}(q^{2})q_{\mu }\right) \,,  \label{ff}
\end{eqnarray}
where $P=p_{1}+p_{2}$ and $q=p_{1}-p_{2}$. The correspondences between our
notations and those used in the literature can be found in the Appendix of
Ref. \cite{CQ2}. The transition amplitude for $B\rightarrow K^{*}l^{+}l^{-}$
is then obtained to be
\begin{equation}
{\cal M}_{K^{*}}^{(\lambda )}=\frac{G_{F}\alpha _{em}V_{tb}V_{ts}^{*}}{2%
\sqrt{2}\pi }\left\{ {\cal M}_{1\mu }^{(\lambda )}L^{\mu }+{\cal M}_{2\mu
}^{(\lambda )}L^{5\mu }\right\}  \label{ampk*}
\end{equation}
with
${\cal M}_{1(2)\mu }^{(\lambda )} =ih_1(g_1)\varepsilon _{\mu \nu \alpha
\beta
}\epsilon ^{*\nu }(\lambda )P^{\alpha }q^{\beta }+h_2(g_2)\epsilon _{\mu
}^{*}(\lambda )+h_3(g_3)\epsilon ^{*}\cdot qP_{\mu }$
where
\begin{eqnarray}
h_{1} &=&C_{9}^{eff}(\mu )V(q^{2})-\frac{2m_{b}}{q^{2}}C_{7}^{eff}(\mu
)T(q^{2}),  \nonumber \\
h_{2} &=&-C_{9}^{eff}(\mu )A_{0}(q^{2})+\frac{2m_{b}}{q^{2}}C_{7}^{eff}(\mu
)T_{0}(q^{2}),  \nonumber \\
h_{3} &=&-C_{9}^{eff}(\mu )A_{1}(q^{2})+\frac{2m_{b}}{q^{2}}C_{7}^{eff}(\mu
)T_{1}(q^{2}),  \nonumber \\
g_{1} &=&C_{10}V(q^{2})\,,\ \
g_{2} \;=\;-C_{10}A_{0}(q^{2})\,,\ \
g_{3} \;=\;-C_{10}A_{1}(q^{2}).  \label{ampk1}
\end{eqnarray}

To have a non-zero T-odd observable, the term of
$\varepsilon _{\mu \nu \alpha \beta }q^{\mu }\epsilon ^{*\nu
}(\lambda )p_{l}^{\alpha }P^{\beta }$ is needed. To get this, we
have to study the processes of $B\rightarrow
K^{*}l^{+}l^{-}\rightarrow (K\pi )l^{+}l^{-}$ so that the
polarizations $\lambda $ and $\lambda ^{\prime }$ in the
differential decay rate, written as $d\Gamma \propto H(\lambda
,\lambda ^{\prime })$ ${\cal M}_{K^{*}}^{(\lambda )}$ ${\cal
M}_{K^{*}}^{(\lambda ^{\prime })\dagger }$ with $H(\lambda
,\lambda ^{\prime })\equiv \epsilon (\lambda )\cdot p_{K}$
$\epsilon ^{*}(\lambda ^{\prime })\cdot p_{K}$, can be different.
 From Eq. (\ref{ampk*}), we see that ${\cal M}_{2\mu }^{(\lambda )} $
only depends on $C_{10}$. Clearly, the T violating effects can not
be generated from ${\cal M}_{2\mu }^{(\lambda )}{\cal M}_{2\mu
^{\prime }}^{(\lambda ^{\prime })\dagger }$, but induced from
${\cal M}_{1\mu }^{(\lambda )}{\cal M}_{1\mu ^{\prime }}^{(\lambda
^{\prime })\dagger }$ and ${\cal M}_{1\mu }^{(\lambda )}{\cal
M}_{2\mu ^{\prime }}^{(\lambda ^{\prime })\dagger }$. This can be
understood as follows. For the ${\cal M}_{1\mu }^{(\lambda )}{\cal
M}_{1\mu ^{\prime }}^{(\lambda ^{\prime })\dagger }TrL^{\mu
}L^{\mu ^{\prime }}$ contribution,
the relevant T-odd terms can
be roughly expressed by
\begin{eqnarray}
{\cal M}_{1\mu }^{(\lambda )}{\cal M}_{1\mu ^{\prime }}^{(\lambda ^{\prime
})\dagger }TrL^{\mu }L^{\mu ^{\prime }} &\propto &Z_{1}{Im}
h_{1}h_{3}^{*}\epsilon (0)\cdot q\varepsilon _{\mu \nu \alpha \beta }q^{\mu
}\epsilon ^{*\nu }(\pm )p_{l^{+}}^{\alpha }P^{\beta }  \nonumber \\
&&+Z_{2}{Im}h_{1}h_{2}^{*}\epsilon (0)\cdot p_{l^{+}}\varepsilon _{\mu \nu
\alpha \beta }q^{\mu }\epsilon ^{*\nu }(\pm )p_{l^{+}}^{\alpha }P^{\beta }
\nonumber \\
&&+Z_{3}{Im}h_{1}h_{2}^{*}\epsilon (\mp )\cdot p_{l^{+}}\varepsilon _{\mu
\nu \alpha \beta }q^{\mu }\epsilon ^{*\nu }(\pm )p_{l^{+}}^{\alpha }P^{\beta
}  \label{im1}
\end{eqnarray}
where $Z_{i}\ (i=1,2,3)$ are functions of kinematic variables and
independent of $C_{9}^{eff}$ and $C_{7}^{eff}$. From Eq. (\ref{ampk1}), one
gets ${Im}h_{1}h_{2}^{*}\sim {Im}h_{1}h_{3}^{*}\sim {Im}C_{9}^{eff}(\mu
)C_{7}^{eff}(\mu )$. We note that as shown in Eq. (\ref{im1}), the T-odd
observables can be non-zero if the process involves a strong phase or
absorptive part even without CP violating phases.
Since both $C_{7,9}^{eff}(\mu )$ include
the absorptive parts, the terms in Eq. (\ref{im1}) do not
vanish in the SM. For ${\cal M}_{1\mu }^{(\lambda )}{\cal M}_{2\mu
^{\prime
}}^{(\lambda ^{\prime })\dagger }TrL^{\mu }L^{5\mu ^{\prime }}$, one gets
\begin{equation}
\left( {\cal M}_{1\mu }^{(\lambda )}{\cal M}_{2\mu ^{\prime }}^{(\lambda
^{\prime })\dagger }+{\cal M}_{2\mu }^{(\lambda )}{\cal M}_{1\mu ^{\prime
}}^{(\lambda ^{\prime })\dagger }\right) TrL^{\mu }L^{5\mu ^{\prime
}}\propto \left( {Im}h_{2}g_{3}^{*}-{Im}h_{3}g_{2}^{*}\right) \varepsilon
_{\mu \nu \alpha \beta }q^{\mu }\epsilon ^{*\nu }(\pm )p_{l^{+}}^{\alpha
}P^{\beta }  \label{im2}\,.
\end{equation}
 From Eq. (\ref
{ampk1}), we find that ${Im}h_{2}g_{3}^{*}-{Im}h_{3}g_{2}^{*}$ is only
related to ${Im}C_{7}^{eff}(\mu )C_{10}^{*}$ and the dependence of ${Im}%
C_{9}^{eff}(\mu )C_{10}^{*}$ is canceled in Eq. (\ref{im2}).
 From Eq. (\ref{effc7}), we see that a nonzero value of
${Im}C_{7}^{eff}(\mu )C_{10}^{*}$ in the SM is an indication of the pure
nonfactorizable effect.

\begin{figure}[tbp]
\vspace{4cm}\centerline{\ \psfig{figure=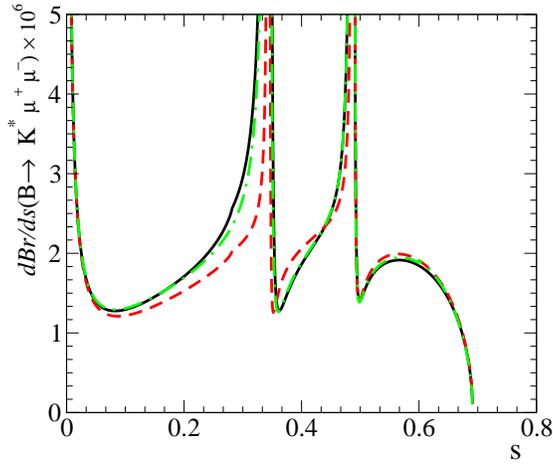,height=2.4in }
}\caption{
 BR of $B\to K^*\mu^+\mu^-$
as a function of $s=q^2/m^{2}_{B}$.
Legend is the same as Figure 1.
} \label{f2}
\end{figure}

In order to write the differential decay rate with the $K^{*}$
polarization, we choose
$\epsilon (0) =\frac{1}{m_{K^{*}}}(|\vec{p}_{K^{*}}|
,0,0,E_{K^{*}})$,
$\epsilon (\pm ) =\left( 0,1,\pm i,0\right)/\sqrt{2}$, and
$p_{l^{+}} =\frac{\sqrt{q^{2}}}{2}( 1,\sin \theta _{l},0,\cos
\theta_{l})$
with
$E_{K^{*}} =(m_{B}^{2}-m_{K^{*}}^{2}-q^{2})/2\sqrt{q^2}$
and
$|\vec{p}_{K^{*}}| =\sqrt{E_{K^{*}}^{2}-m_{K^{*}}^{2}}$
in the $q^{2}$ rest frame and $p_{K}= (1,\sin \theta _{K}\cos \phi,\sin
\theta _{K}\sin \phi ,\cos \theta _{K}) m_{K^{*}}/2$ in the $K^{*}$ rest
frame where $\phi $ denotes the relative angle of the decaying plane between
$K\pi $ and $l^{+}l^{-}$.
We have that

\begin{eqnarray}
&&\frac{d\Gamma }{d\cos \theta _{K}d\cos \theta _{l}d\phi dq^{2}}=\frac{%
3\alpha _{em}^{2}G_{F}^{2}\left| \lambda _{t}\right| ^{2}\left| \vec{p}%
\right| }{2^{14}\pi ^{6}m_{B}^{2}}Br(K^{*}\rightarrow K\pi )  \nonumber \\
&&\ \ \times \left\{ 4\cos ^{2}\theta _{K}\sin ^{2}\theta
_{l}\sum_{i=1,2}\left| {\cal M}_{i}^{0}\right| ^{2}+\sin ^{2}\theta
_{K}(1+\cos ^{2}\theta _{l})\sum_{i=1,2}\left( \left| {\cal M}%
_{i}^{+}\right| ^{2}+\left| {\cal M}_{i}^{-}\right| ^{2}\right) \right.
\nonumber \\
&&-\sin 2\theta _{K}\sin 2\theta _{l}\sin \phi \sum_{i=1,2}{Im}\left( {\cal M%
}_{i}^{+}-{\cal M}_{i}^{-}\right) {\cal M}_{i}^{0*}-2\sin ^{2}\theta
_{K}\sin ^{2}\theta _{l}\sin 2\phi \sum_{i=1,2}{Im}\left( {\cal M}_{i}^{+}%
{\cal M}_{i}^{-*}\right)  \nonumber \\
&&\left. +2\sin 2\theta _{K}\sin \theta _{l}\sin \phi \left( {Im}{\cal M}%
_{1}^{0}({\cal M}_{2}^{+*}+{\cal M}_{2}^{-*})-{Im}({\cal M}_{1}^{+}+{\cal M}%
_{1}^{-}){\cal M}_{2}^{0*}\right) \,+...\right\} ,  \label{angular}
\end{eqnarray}
where
$ |\vec{p}| =\sqrt{(m_B^2+m_{K^{*}}^2-q^2)^2/4m_b^2-m_{K^{*}}^{2}}$
and ${\cal M}_{i}^{0,\pm }$
denote the longitudinal and transverse polarizations of $K^{*}$, and their
explicit expressions are given by
\begin{eqnarray*}
{\cal M}_{a}^{0} =\sqrt{q^{2}}\left(
\frac{E_{K^{*}}}{m_{K^{*}}}f_{2}+2%
\sqrt{q^{2}}\frac{\left| \vec{p}_{K^{*}}\right| ^{2}}{m_{K^{*}}}f_{3}\right)
\ {\rm and}\
{\cal M}_{a}^{\pm } =\sqrt{q^{2}}\left( \pm 2\left| \vec{p}_{K^{*}}\right|
\sqrt{q^{2}}f_{1}+f_{2}\right)\,,
\end{eqnarray*}
respectively, where $a=1(2)$ while $f=h(g)$. For simplicity, we just show
the relevant terms in Eq. (\ref{angular}). The detailed derivation will be
discussed elsewhere \cite{CQ3}. Other distributions for the $K^{*}$
polarization and CP violating observables
can be found in Refs. \cite{Kim,Kruger,Grossman}. From Eqs. (\ref{im1})
and (\ref
{im2}), we know that ${Im}({\cal M}_{i}^{+}-{\cal M}_{i}^{-}){\cal M}%
_{i}^{0*}$ and ${Im}({\cal M}_{i}^{+}{\cal M}_{i}^{-*})$ are from ${\cal M}%
_{1\mu }^{(\lambda )}{\cal M}_{1\mu ^{\prime }}^{(\lambda ^{\prime })\dagger
}TrL^{\mu }L^{\mu ^{\prime }}$ while ${Im}{\cal M}_{1}^{0}({\cal M}_{2}^{+*}+%
{\cal M}_{2}^{-*})-{Im}({\cal M}_{1}^{+}+{\cal M}_{1}^{-}){\cal M}_{2}^{0*}$
is induced by ${\cal M}_{1\mu }^{(\lambda )}{\cal M}_{2\mu ^{\prime
}}^{(\lambda ^{\prime })\dagger }TrL^{\mu }L^{5\mu ^{\prime }}$.
 In Figure \ref{f2}, we show the effect of the various
parametrizations on the differential decay rate
after integrating over angles in Eq. (\ref{angular}).
 As seen from the
figure, there are not many differences among the three scenarios
except the result in (II) with the LD effect. Obviously, by
measuring the decay rate, one could not be able to tell which
scenario of the LD parametrizations is favorable.

In order to explore the
possibility of extracting LD effects, we examine the observables,
defined by
\begin{equation}
\langle {\cal O}_{i}\rangle \equiv \int {\cal O}_{i}{\frac{d\Gamma }{dq^{2}}}
\label{avgO}
\end{equation}
where ${\cal O}_{i}$ are momentum correlation operators, given by
\begin{eqnarray}
{\cal O}_{L} &=&4\frac{\left| \vec{p}_{l^{+}}\times \vec{p}_{B}\right| ^{2}}{%
\left| \vec{p}_{B}\right| ^{2}\omega _{l^{+}}^{2}}-3\frac{\left| \vec{p}%
_{B}\times \vec{p}_{K}\right| ^{2}}{\left| \vec{p}_{B}\right| ^{2}\omega
_{K}^{2}} \\
{\cal O}_{T1} &=&\frac{\left( \vec{p}_{B}\cdot \vec{p}_{l^{+}}\times \vec{p}%
_{K}\right) \left( \vec{p}_{B}\times \vec{p}_{K}\right) \cdot \left( \vec{p}%
_{l^{+}}\times \vec{p}_{B}\right) }{\left| \vec{p}_{B}\right| ^{3}\omega
_{K}^{2}\omega _{l^{+}}^{2}} \\
{\cal O}_{T2} &=&\frac{\left( \vec{p}_{B}\cdot \vec{p}_{K}\right) \left(
\vec{p}_{B}\cdot \vec{p}_{l^{+}}\times \vec{p}_{K}\right) }{\left| \vec{p}%
_{B}\right| ^{2}\omega _{K}^{2}\omega _{l^{+}}}
\end{eqnarray}
with $\omega _{K}=m_{K^{*}}/2$ and $\omega _{l^{+}}=\sqrt{q^{2}}/2$. In the $%
K^{*}$ rest frame, we note that ${\cal O}_{L}=4\sin ^{2}\theta _{l}-3\sin
^{2}\theta _{K}$, ${\cal O}_{T1}=\sin ^{2}\theta _{K}\sin ^{2}\theta
_{l}\cos \phi \sin \phi $ and ${\cal O}_{T2}=\cos \theta _{K}\sin \theta
_{K}\sin \theta _{l}\sin \phi $. Explicitly, one has that
\begin{eqnarray}
\langle {\cal O}_{L}\rangle  &\propto &\sum_{i=1,2}\left| {\cal M}%
_{i}^{0}\right| ^{2}\,,  \nonumber \\
\langle {\cal O}_{T1}\rangle  &\propto &\sum_{i=1,2}Im({\cal M}_{i}^{+}{\cal %
M}_{i}^{-*})\,, \nonumber \\
\langle {\cal O}_{T2}\rangle  &\propto &Im{\cal M}_{1}^{0}({\cal M}_{2}^{+*}+%
{\cal M}_{2}^{-*})-Im({\cal M}_{1}^{+}+{\cal M}_{1}^{-}){\cal
M}_{2}^{0*}\,. \label{operators}
\end{eqnarray}
We note that the result from the first T-even (odd) term in Eq.
(\ref{angular}) is similar to that from the second one.
As shown in Eqs. (\ref{im1}) and (\ref{im2}),
the T-odd observables of $\langle {\cal O}_{T1,T2}\rangle$
 in  Eq. (\ref{operators}) are related to ${Im}C_{9}^{eff}C_{7}^{eff*}$
and ${Im}C_{7}^{eff}C_{10}^{*}$, respectively.
The
statistical significances of the observables in Eq. (\ref{avgO})
can be determined by
\begin{equation}
\varepsilon _{i}(q^{2})={\frac{\int {\cal {O}}_{i}{\frac{d\Gamma }{dq^{2}}}}{%
\sqrt{\int {\frac{d\Gamma }{dq^{2}}}\int {\cal O}_{i}^{2}{\frac{d\Gamma }{%
dq^{2}}}}}}\,.  \label{epsilon}
\end{equation}

\begin{figure}[tbp]
\centerline{ \psfig{figure=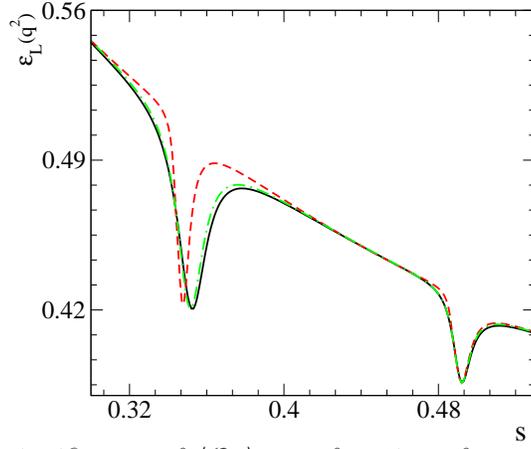,height=2.4in }} \caption{
 The statistical
significance of $\langle {\cal O}_{L}\rangle$ as a function of
$s=q^2/m^{2}_{B}$. Legend is the same as Figure 1.
 } \label{f3}
\end{figure}

\begin{figure}[tbp]
\vspace{4cm}\centerline{ \psfig{figure=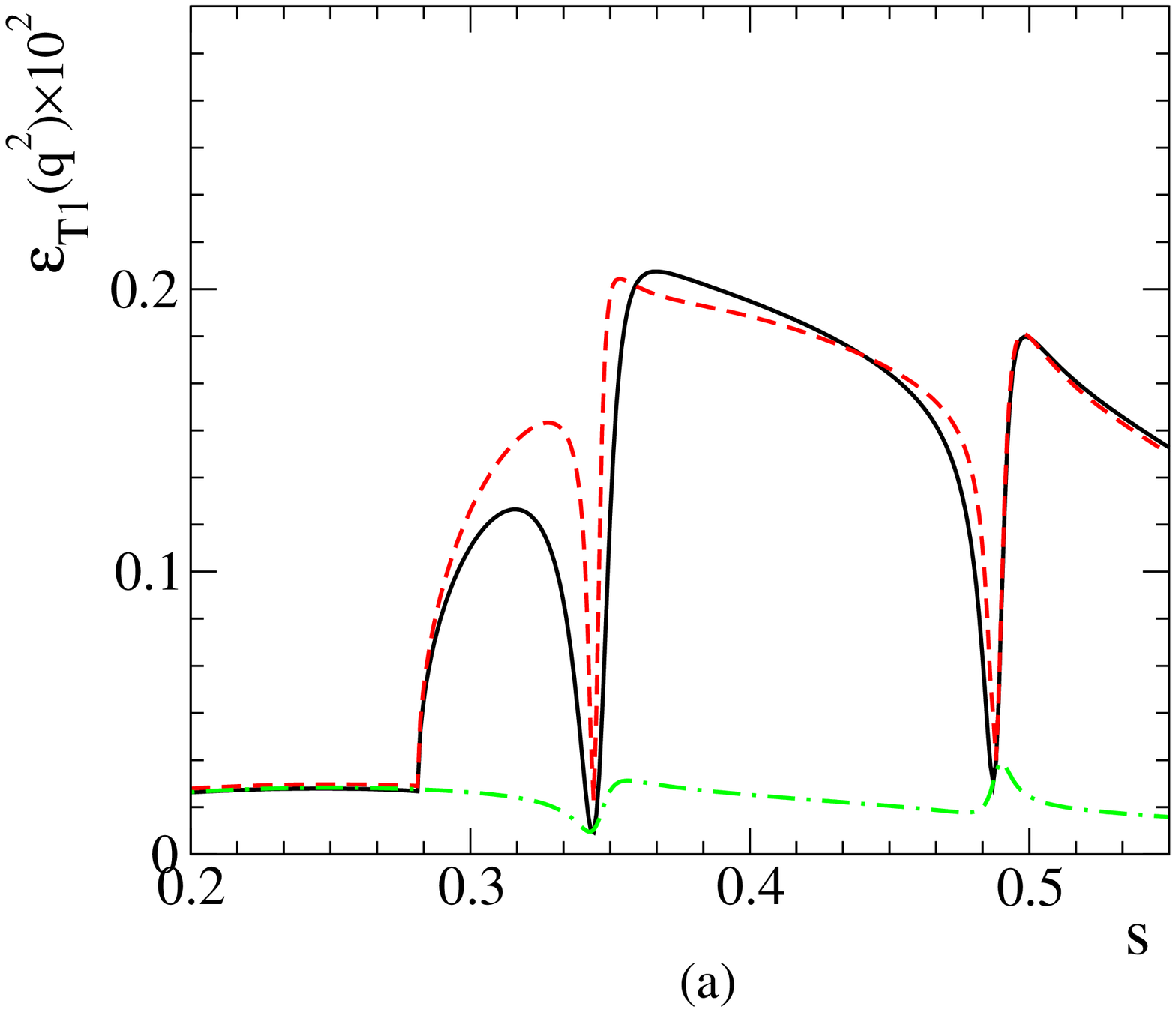,height=2.4in
}$\ \ $ \psfig{figure=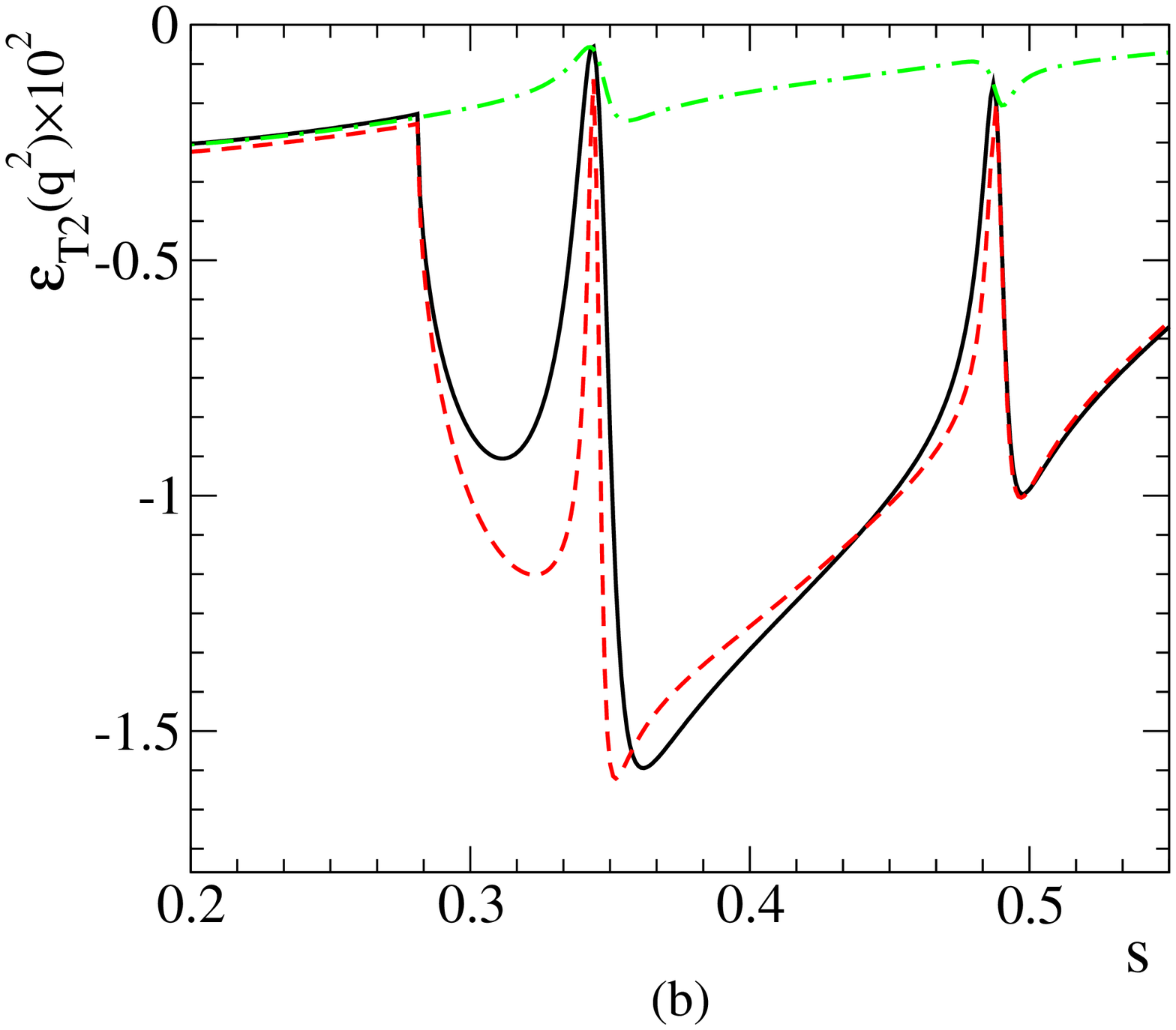,height=2.4in }} \caption{Same
as Figure \ref{f3} but for (a) $\langle {\cal O}_{T1}\rangle $ and
(b) $\langle {\cal O}_{T2}\rangle$ with $\omega=-0.15$. }
\label{f4}
\end{figure}

In Figures \ref{f3} and \ref{f4}, we show the statistical
significances for ${\cal O}_{T1,T2}$ as functions of $s$
for various cases. From these figures, we see that: (a) the
effects on the T-even observable of $\langle O_{L}\rangle$ are
large and the contributions to $ \varepsilon _{L}$ from  scenarios
(I) and (III) are slight different from (II) around the first
resonance region; (b) the contributions in the scenario (III) to
the T-odd observables of $\langle O_{T1,T2}\rangle$ are much
smaller than the other two scenarios and those in (I) and (II) are
almost the same except the region close to the first resonance;
and (c) the effects of LD contributions to $\varepsilon_{T1}$ are
much less than $1\%$ but those to $\varepsilon_{T2}$ are at the
percent level. It is interesting to note that the differences on
the results of $\langle O_{T1(2)}\rangle$ between (I,II) and (III)
are significant.
Moreover, it is worth to emphasize that the results of Figure
\ref{f4} are purely from nonfactorizable contributions. For
example, in the SM, a signal of $\varepsilon_{T2}$ will directly
reflect the nonfactorized effects.

In summary, we have defined several momentum correlations as
physical observables in $B\rightarrow K^{*}l^{+}l^{-}$ decays with
the polarized $ K^{*}$  to study the LD contributions in the SM.
we have found that these observables are quite sensitive to the
different scenarios of the LD parametrizations. In particular, we
have illustrated that the nonfactorizable effect of $B\to J/\Psi
K^*$ for the T-odd observable of $\langle O_{T2}\rangle$ is
non-negligible.
Searching for $\langle O_{T2}\rangle$ could distinguish various
parametrizations of the LD contributions in exclusive heavy $B$ meson
decays.
Finally, we remark that if there is new physics beyond the SM,
such as the leptoquark and supersymmtric models, our results here
can be treated as theoretical backgrounds and  the new physics
contributions to observables are easily at the level of $10\%$
\cite{CQ3}.
\newline

\noindent {\bf Acknowledgments}

This work was supported in part by the National Science Council of the
Republic of China under Contract Nos. NSC-90-2112-M-001-069 and
NSC-90-2112-M-007-040.


\newpage

\ed